\documentclass[pra,twocolumn,a4paper]{revtex4}

\usepackage{amsmath}
\usepackage{amssymb}
\usepackage{graphicx}

\begin{document}
\author{Thomas Salzburger and Helmut Ritsch}
\title{Twin stimulated amplification of light and matter waves in an atom--photon pair laser}
\affiliation{Institute for Theoretical Physics, University Innsbruck\\ Technikerstrasse 25/2, 6020 Innsbruck, Austria}
\date{\today}

 \begin{abstract}
We consider ultracold atoms in a far detuned optical lattice orientated across a high-$Q$ optical resonator. Applying an external driving laser to the atoms, which is red detuned from the cavity mode by one vibrational quantum, induces cavity-enhanced sideband cooling. For a dense and cold enough atomic ensemble we predict an oscillation threshold for optical Raman sideband lasing concurrent with coherent matter-wave amplification. Above this oscillation threshold photons and atoms in the lowest band are dominantly created pairwise via stimulated emission with a strong suppression of competing spontaneous processes. In close analogy to a nondegenerate parametric oscillator we find sub-Poissonian photon statistics and almost perfect nonclassical atom--photon number correlations. Injecting atoms in higher vibrational bands via tunneling or incoherent scattering then leads to continuous, simultaneous generation of a coherent atom beam and laser light with nonclassical atom--field correlations.        
\end{abstract}

\maketitle

\section{Introduction}
Already a century ago the striking analogy between atomic Schr\"odinger waves and electromagnetic waves has started a new field in physics studying wave properties of massive particles \cite{DeBroglie24}. Nevertheless, only the significant advances in the cooling and trapping techniques of neutral atoms in the last decades \cite{Nobel1} made extensive and well controlled experimental investigations of macroscopic coherent matter waves possible \cite{CoolingNobel}.

A central and ongoing quest in this field is the realization of the atomic counterpart of an optical laser generating a coherent intense matter-wave beam. Although first proposals to implement such an atom laser suggested the use of refined optical cooling techniques within optical traps \cite{Wiseman95,Spreeuw95,Olshanii95}, it was the creation of Bose-Einstein condensates via evaporative cooling which paved the way to coherent atomic beam sources by output coupling from such a condensate \cite{Mewes97, Anderson98,Hagley99,Bloch99}. Plenty of follow-up experiments based on this principle have been performed since, including the study of the spatial \cite{Bloch00} as well as temporal \cite{Kohl01} coherence, the divergence \cite{Coq01}, and the counting statistics \cite{Ottl05} of the produced atomic beams.

Despite some improvements a quasi continuous-wave (cw) implementation of an atomic beam out-coupled from a pre-existing Bose-Einstein condensate (BEC) is still hampered by the finite number of atoms initially present in the condensate. This limits the pulse duration and the phase and intensity stability of the matter-wave beam. The central challenge to overcome here is continuous reloading of atoms. Naturally a pump source is needed for a cw operation of a laser. As atoms cannot be created out of nothing, one cannot just inject energy itself but the only way to do this for an atom laser is to feed new atoms into the BEC mode \cite{Ketterle02,Pfau02}.

In a straightforward but technologically challenging approach, one has tried to convert the time sequence of BEC generation into a linear array of spatially separated cooling stages to operate a BEC generator in the continuous regime \cite{Dalibard04}. These efforts already produced a very cold and partially coherent atomic beam, but to achieve full atom lasing, the setup still needs further improvements. 

Theoretically, several alternative proposals for a constantly operating atom laser by continuous atomic out-coupling from a BEC were put forward including continuous injection of cold atoms and ongoing evaporative cooling \cite{Holland96,Guzman96,Wiseman96}.
The re-filling of condensate atoms is then achieved by a collisional transfer of atoms into the BEC mode out of a large reservoir.
While the corresponding microscopic implementation of the injection mechanism may be manifold, the temperature of the newly injected atoms typically has to be close to the critical value for condensation. When using near-resonant lasers for the required optical pre-cooling during the evaporative cooling process, one again recovers the problem of reabsorption heating \cite{Castin98} as only high phase-space densities can assure the required atomic collision rates for fast enough replenishing. These problems are then closely related to the difficulties appearing in the early proposals for atom lasing via all optical cooling \cite{Wiseman95,Spreeuw95,Olshanii95,Santos2001}. 

In this paper we study a different scenario for continuous optical replenishing the BEC mode at much lower density by help of an optical resonator to tailor the optical atomic cooling process \cite{Salzburger07}. Such a cavity helps to suppress spontaneous emission and thus indirectly prevents reabsorption. Naturally cavity cooling in its standard form was already considered as a tool to achieve a BEC by all optical means \cite{Horak01}. In general the final temperature is only limited by the cavity linewidth, and photons emitted through the mirrors cannot be reabsorbed. Hence, this technology should be in principle sufficient to achieve degeneracy, if a good enough cavity can be build around a dense atomic cloud. Such a very narrow cavity line requires however a challenging technological setup, and the timescales involved in the dynamics get rather slow.

Here we study an improved and conceptually simpler variant of a cavity-based setup for an atom laser, which is designed to work under different and technologically less stringent conditions. The basic idea is to implement a combination of an atom laser and a photon laser operating simultaneously on a vibrational anti-Stokes Raman transition. Here the central physical mechanism to suppress the unwanted transitions and photon emissions brought about by the cavity is not strong atom--field coupling but stimulated enhancement of the desired transition. This allows to suppress reabsorption as well, but still implies fast transition times. Note that a related setup was suggested to enhance sideband cooling in ion traps \cite{Cirac95,Zippilli05}. Instead of a single trapped ion, here we consider a large ensemble of ultracold atoms in a far detuned optical lattice and the case of a good cavity.

The occupation of the lowest lattice states is then increased by a stimulated transfer from a higher vibrational excitation (or a band in a lattice). Actually, for a single trapped atom, coherent coupling of atomic quantum motion and a field mode in cavities via stimulated Raman transitions has been suggested and investigated already in earlier work \cite{Parkins99}. In an anti-Stokes transition, even the vacuum field can stimulate the transfer of an atom into a vibrational lower state, simultaneous with induced photon emission into a cavity mode. When the photon leaves the resonator no backward process is possible.

For a larger ensemble of $N$ atoms this single photon emission process is collectively enhanced by a factor $N$ with respect to a spontaneous transition \cite{Beige05}. Entering the regime where the Raman gain on the sideband transition exceeds the cavity losses, more than a single photon is present and lasing via stimulated emission will start. In this case one can make use of double Bose enhancement through macroscopic occupation of the atomic mode and the cavity mode. Obviously the total transition rate is then proportional to the total photon emission through the cavity. 

Including replenishing of atoms to higher vibrational lattice levels and out-coupling of atoms from the lowest band should then lead to continuous lasing and atom beam emission. The goal of the present paper is to develop a consistent theoretical description of the essential physics of this pairwise stimulated enhancement and to study the nature of the generated output beams of light and atoms in more detail as already presented in our short letter \cite{Salzburger07}. Due to the complexity of a practical implementation we have to refrain to a series of simplifying assumptions to derive a solvable model as outlined below. 

From a different perspective the present paper can also be viewed as a continuation of the work on cavity-cooling of atoms induced by stimulated emission \cite{Salzburger04,Salzburger05,Salzburger06}. There we investigated the light forces in a laser system composed of a single inverted atom \cite{Salzburger04,Salzburger05} or even an inverted atomic ensemble \cite{Salzburger06} in a high-$Q$ cavity. We found that the optical gain through stimulated emission in the system can be tied to motional cooling and atomic self-trapping. In this system several effects including lasing, trapping, and cooling work together in an appropriate way such that atomic temperatures well below the limit of passive cavity-cooling are possible \cite{Salzburger05,Salzburger06}. Nevertheless the system stays far away from degeneracy as the atomic inversion is accompanied by noise and spontaneous emission. Here by help of an external potential and employing a vibrational Raman transition, we can relax the condition of internally inverted atoms. In addition, the external lattice provides for trapping so that the light force induced by the laser light are not needed for trapping, and thus significantly less dipole heating is present.

It is well established theoretically and practically that sideband cooling of a single atom (ion) efficiently reaches the ground state. Theoretical predictions suggest that this behavior can be further improved utilizing a cavity \cite{Cirac95,Zippilli05,Murr06}. While in the free-space case generalization to many particles is a hard task, collective enhancement should even be helpful in the cavity-enhanced case \cite{Beige05}. 

The central remaining problem still is reloading of new atoms without disturbing the Raman gain process. There seem to be several solutions to this process. As a first suggestion the lasers applied for creating a lattice or a trap within the cavity can also be tailored to guide extra atoms from a pre-cooled reservoir into the active lasing zone as experimentally demonstrated \cite{Nussmann05,Kuhr,Sauer04,Yang07}. This can happen in quite different forms. When lasers are applied to form, e.g., a quasi 1D optical lattice potential along a standing wave laser beam, the guided atoms can simply enter and leave the cooling region on the basis of tunneling \cite{Wiseman95}. In another possibility the laser directly involved in the final cooling stage to reach degeneracy by pumping atoms from a cold reservoir into the lasing modes could be used \cite{Spreeuw95,Moy97}. This even avoids the need of spatial separation of the pre-cooling and the lasing stages.

\begin{figure}
  \includegraphics[width=8cm]{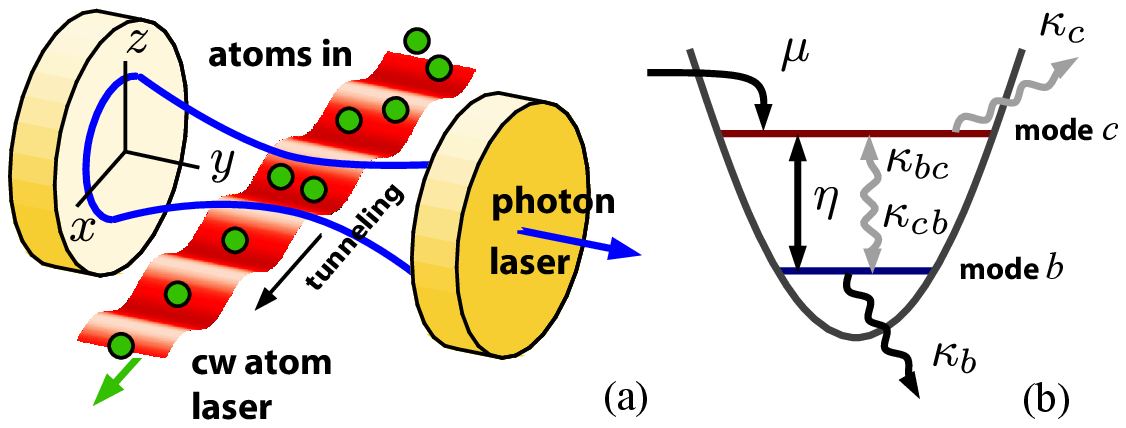}
  \caption{(a) Illustration of the cavity--lattice configuration forming the atom--photon pair laser. Both light and matter fields are stimulatingly amplified by Raman-gain involving blue motional sidebands.\\
  (b) Illustration of the atomic motional dynamics including injection ($\mu$), cooling ($\eta$), and out-coupling ($\kappa_b$), as well as losses due to heating ($\kappa_{bc},\kappa_{cb},\kappa_{c}$).}
  \label{fig0}
\end{figure}

Here, as sketched in Fig.\ \ref{fig0}(a), we consider a quite generic scenario where several of the above possibilities are included. Atom loading from a reservoir into the trap potential is simply modeled by effective rates, which can be coherent as for tunneling or incoherent via scattering. Our work is organized as follows:\ In Sec.~\ref{model} we give an overview of our atom--photon laser scheme. In Sec.~\ref{rates}, we model it by an incoherent transfer of atoms into the trap as it could be realized, e.g., from a thermal source \cite{Holland96} or by an incoming cold atomic beam \cite{Wiseman96}. On the other hand, we use in Sec.~\ref{sec:tunnel} a model based on atomic tunneling through the lattice which we already used in \cite{Salzburger07}. In Sec.~\ref{rates} we use a rate-equation approach for the excitation probabilities of the pair laser and study mean number and number statistics in the steady state. An extended model description based on stochastic differential equations is derived in Sec.~\ref{sec:tunnel} which allows us to investigate the noise properties and the spectra of the output beams in Sec.~\ref{output}.

\section{Model}
\label{model}
Let us now present a mathematical model for a system as shown in Fig.\ \ref{fig0}, which exhibits the mechanisms underlying an atom--photon twin laser in a rather generic form, omitting as much specific experimental details as possible. The basic components are a large ensemble of cold two-level atoms placed in an external optical trap which is intersected by the field of an optical resonator with high quality. We consider the simple situation where the atomic trap consists of an optical lattice across the resonator. At each lattice site we assume at least two well separated vibrational energy levels. The manifolds of eigenstates of the lattice sites within the cavity are identified with the atomic source and lasing modes. In the simplest case of a long-wavelength lattice, only a single lattice site might be located within the resonator mode volume. Analogous to cavity enhanced sideband cooling \cite{Cirac95}, we then tune an external pump laser and the cavity mode frequency to Raman resonance between these two trap states, which resonantly couples the cavity photons and the atomic (quantized) motional states \cite{Parkins99}. We choose the frequencies in a way that an atom is transferred from the upper (source) to the lower (ground) vibrational level whenever a photon blue detuned to the pump is generated in the cavity mode.

For large enough detuning from the upper Raman level, the only significant dissipation channel in this system is cavity decay. In this limit each photon emitted through the cavity mirrors increases the occupation number in the lower atomic state by one. The process in its ideal form only stops when no more atoms are present in the upper level \cite{Beige05}. The central important point to note here is that the Raman coupling and thus the rate for the pair creation of photons and ground-state atoms is collectively enhanced with growing cavity-photon number and number of atoms already present in the ground level. Hence even if it is weak for a single atom, for increasing numbers of atoms and photons this desired process will more and more dominate other, unwanted transitions in the system.

Note that in a realistic three dimensional configuration one probably will have several lattice wells within the cavity mode. Nevertheless, as long as collisions or atomic on-site interactions are weak, we can formally combine them into a single effectively trapping state or mini band. Hence we effectively replace the lattice bands simply by two effective levels (mini bands) describing the atomic source and laser modes \cite{Salzburger07}.

The free Hamiltonian is just represented by three oscillators in the form
\begin{equation}
  H_\mathrm{free}=\hbar\omega_\mathrm{a}a^\dagger a+E_0b^\dagger b+E_1c^\dagger c.
  \label{free}
\end{equation}
The operators $a$, $b$, and $c$ represent the cavity photons and the two atomic modes describing the atom laser and atom source states, respectively; $\omega_\mathrm{a}$ denotes the bare cavity resonance frequency; $E_0$ and $E_1$ are the energies of the atomic bands. In a first approximation the Raman transition can then be described by an effective nonlinear three-mode coupling Hamiltonian of the form \cite{Zeng94}
\begin{equation}
  H_\mathrm{int}=i\hbar\eta\left(a^\dagger b^\dagger c-a\,b\,c^\dagger\right)
  \label{pair}
\end{equation}
where $\eta$ is the interaction strength. Note that with respect to our later goal of atom lasing, we have already dropped the non-resonant heating terms. For a single particle this is only valid for sufficient trap depth \cite{Zippilli05}. A more realistic approach has to contain off-resonant transitions as, e.g., discussed in Ref.~\cite{Beige05}.

In order to allow for steady-state operation, we have to include losses from the laser mode. Possible such coherent output-coupling processes include, e.g., atomic tunneling \cite{Wiseman95} or the application of a Raman transition for an atomic state change into an untrapped state \cite{Moy97,Ottl05}. Mathematically, independent of the specific implementation, the out-coupling mechanism is treated via a linear loss term in the master equation in Lindblad form analogous to photon loss. The master equation including cavity decay reads thus
\begin{align}
  \label{rho}
  \dot{\rho}&=-i/\hbar\left[H_\mathrm{int},\rho\right]+\mathcal{L}\rho,\\
  \mathcal{L}\rho&=\kappa_a\left(\left[a,\rho\,a^\dagger\right]+\left[a\rho,a^\dagger\right]\right)\nonumber\\
  \label{lrho}
  &+\kappa_b\left(\left[b,\rho\,b^\dagger\right]+\left[b\rho,b^\dagger\right]\right),
\end{align}
where $\kappa_a$ and $\kappa_b$ denote the cavity (amplitude) decay and atomic out-coupling strength, respectively.

In addition to these terms we will also include spontaneous Raman transitions from the upper to the lower atomic states without generating a cavity photon occurring at rate $\kappa_{cb}$, which will be important only below threshold but are included for completeness here.
The corresponding term $\mathcal{L}_{cb}$ in the master equation reads
\begin{equation}
  \mathcal{L}_{cb}=\kappa_{cb}\left(\left[b^\dagger c,\rho\,c^\dagger b\right]+ \left[b^\dagger c\,\rho,c^\dagger b\right]\right).
  \label{spontraman}
\end{equation}
Note that losses induced by the Raman pump laser on the atomic ground and upper states can be simply included into effective corresponding loss rates $(\kappa_b,\kappa_c)$ and will be small for a far detuned Raman transition. 

A central and critical point in the model is of course the refilling of the atoms in the source mode, which is required for a steady laser operation. As the size of the full lattice is typically orders of magnitudes larger than the cavity volume, the trapped atoms outside the mode already form a useful reservoir for this purpose. In principle this reservoir might be even periodically refilled from a nearby molasses operated far detuned from the Raman pump laser. A more decisive challenge is to design a mechanism how these reservoir atoms then can be continuously transferred into the interaction region within the cavity. While various methods for this are conceivable in practice, at the end they can be modeled by only two different generic types of pumping, namely via coherent or incoherent transfer of atoms to the upper source mode. While the first corresponds to a tunneling type injection from filled sites outside the resonator or a deterministic injection by actively moving the filled lattice through the resonator \cite{Kuhr}, the second describes refilling by scattering between atoms or optical pumping from higher lying states.

\begin{figure}[htb]
  \includegraphics[width=6.8cm]{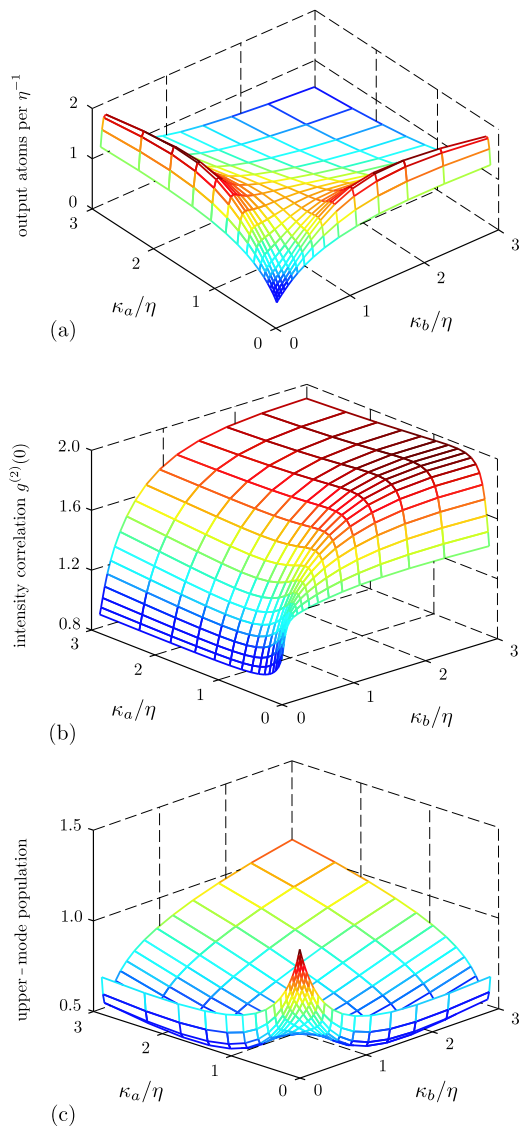}
  \caption{(a) Emission rate of condensed atoms from the lower vibrational band as a function of cavity linewidth $\kappa_a$ and atom out-coupling rate $\kappa_b$ for $\mu_\mathrm{out}=0.9\mu_\mathrm{in}=3\eta/8$ and $\kappa_{cb}=\eta/80$.\\
    (b) Intensity correlation function $g_2(0)$ of the condensed atoms from the lower vibrational band as a function of cavity linewidth $\kappa_a$ and atom out-coupling rate $\kappa_b$ for parameters as above.\\
    (c) Atomic occupation number of the upper (source) vibrational band as a function of cavity linewidth $\kappa_a$ and atom out-coupling rate $\kappa_b$ for parameters as above.}
  \label{fig15}
\end{figure}

Before discussing these two possibilities and their consequences in sections \ref{rates} and \ref{sec:tunnel} in detail, we will first give some qualitative insight into the system via a direct numerical solution of our model, i.e., Eq.\ \eqref{rho}. This is only possible for very low atom and photon numbers, but nevertheless some general features of the dynamics can be already nicely seen here. In Fig.\ \ref{fig15}(a) we plot the intensity of emitted atoms from the lower band as a function of the decay rate of the optical resonator ($\kappa_a$) and the atom out-coupling rate ($\kappa_b$) for fixed atom-field coupling $\eta$. The system is assumed to reach a stationary state due to atomic replenishing as described in more detail in the next section.
We clearly see that a high-$Q$ optical cavity enhances the atom emission almost as much as a low atomic loss rate. This is directly related to the dominance of stimulated pairwise atom-photon generation. The corresponding intensity correlation function (pair correlation) $g_2(0)$ shown in Fig.\ \ref{fig15}(b) clearly illustrates the transition from thermal bunching $g_2(0)=2$ to a Poissonian value $g_2(0) \approx 1 $ as for a laser crossing threshold. Atomic anti-bunching seems possible even without any direct interaction between the atoms.
The stimulated enhancement of the twin gain via atoms and photons is shown in Fig.\ \ref{fig15}(c) where we depict the stationary upper-atomic-state population. Due to this enhancement the upper level gets depleted below one atom on average, and the atoms are preferentially deposited in the atom laser mode before they leak out. 

While the above numerical results confirm the basic idea, these calculations do not come near to experimental realistic atom and photon numbers. Hence in the following we will develop further analytic approximative descriptions to reach more interesting and realistic parameter regimes.

\section{Incoherent injection of atoms}
\label{rates}
We first assume that the injection of atoms is realized by a pre-cooled atomic beam injected into the trap. This will incoherently add and remove occupations from the atomic trap levels. Since the trap acts as a resonator cavity for matter waves, we model this by incoherent occupation and depletion of the individual levels. In principle, we have four individual rates, altogether.
However, as the ground mode is not intended to be directly pumped, we restrict the injection of atoms to the upper atomic (source) mode $c$. We further allow atoms to be lost from mode $c$ due to heating or re-thermalization processes, which might be included into the loss rate for the upper level.
Identifying $\mu_\mathrm{in}$ and $\mu_\mathrm{out}$ as the atomic incoming and outgoing rates, respectively, a corresponding model term in the master equation reads
\begin{align}
  \mathcal{L}_\mathrm{pump}\rho&=\mu_\mathrm{in}\left(\left[c^\dagger,\rho\,c\right]+\left[c^\dagger\rho,c\right]\right)\nonumber\\
  &+\mu_\mathrm{out}\left(\left[c,\rho\,c^\dagger\right]+\left[c\,\rho,c^\dagger\right]\right).
\end{align}

Let us now have a look at the dynamics induced by this model. When there is at most one atom at a time present in the source mode, we can search for a solution in the reduced basis $|n,m,k\rangle$ representing $n$ photons in the cavity mode, $m$ atoms in the ground level, and $k=0$ or $1$ atoms in the source level. The joint probability of having $n$ photons and $m$ atoms is conveniently written down as
\begin{gather}
  \label{jointprobdef}
  P_{n,m}\approx p_{n,m,0}+p_{n,m,1}
\end{gather}
with the diagonal density-matrix elements
\begin{equation}
  \label{pnmks}
  p_{n,m,k}=\langle n,m,k|\rho|n,m,k\rangle.
\end{equation}

The evolution of the joint atom--photon number statistics $P_{n,m}(t)$ is hence determined by a set of coupled differential equations for the probabilities $p_{n,m,0}$ and $p_{n,m,1}$ which can be directly calculated from the master equation. They read
\begin{subequations}
\label{pnms}
\begin{align}
\label{pnm0}
\frac{d}{dt}p_{n,m,0}&=\eta\sqrt{n m}\,V_{n,m}\nonumber\\
  &+2\kappa_a\left[(n+1)p_{n+1,m,0}-n\,p_{n,m,0}\right]\nonumber\\
  &+2\kappa_b\left[(m+1)p_{n,m+1,0}-m\,p_{n,m,0}\right]\nonumber\\
  &-2\mu_\mathrm{in}\,p_{n,m,0}+2\mu_\mathrm{out}\,p_{n,m,1},\\
\label{pnm1}
\frac{d}{dt}p_{n,m,1}&=-\eta\sqrt{(n+1)(m+1)}\,V_{n+1,m+1}\nonumber\\
  &+2\kappa_a\left[(n+1)\,p_{n+1,m,1}-n\,p_{n,m,1}\right]\nonumber\\
  &+2\kappa_b\left[(m+1)\,p_{n,m+1,1}-m\,p_{n,m,1}\right]\nonumber\\
  &+2\mu_\mathrm{in}\,p_{n,m,0}-2\mu_\mathrm{out}\,p_{n,m,1},\\
%\dot{p}_{n-1,m-1,1}&=-\eta\sqrt{n m}\,V_{n,m}\nonumber\\
%  &+2\kappa_a\left[n\,p_{n,m-1,1}-(n-1)\,p_{n-1,m-1,1}\right]\nonumber\\
%  &+2\kappa_b\left[m\,p_{n-1,m,1}-(m-1)\,p_{n-1,m-1,1}\right]\nonumber\\
%  &+2\mu_\mathrm{in}\,p_{n-1,m-1,0}-2\mu_\mathrm{out}\,p_{n-1,m-1,1},\\
\label{vnm}
\frac{d}{dt}V_{n,m}&=-(\mu_\mathrm{in}+\mu_\mathrm{out})V_{n,m}\nonumber\\
  &+2\eta\sqrt{n m}\left(p_{n-1,m-1,1}-p_{n,m,0}\right)\nonumber\\
  &+2\kappa_a\sqrt{n(n+1)}V_{n+1,m}-(2n-1)\kappa_a V_{n,m}\nonumber\\
  &+2\kappa_b\sqrt{m(m+1)}V_{n,m+1}-(2m-1)\kappa_b V_{n,m}
\end{align}
\end{subequations}
where the off-diagonal elements are defined as
\begin{align}
  V_{n,m}&=\langle n-1,m-1,1|\rho|n,m,0\rangle\nonumber\\
    &+\langle n,m,0|\rho|n-1,m-1,1\rangle.
\end{align}
Note that in Eqs.~\eqref{pnms} we have dropped terms involving the dynamics of higher-order probabilities $p_{n,m,i>1}$ to guarantee that the feed of atoms into the source mode has no direct influence on the atom--photon distribution $P_{n,m}(t)$. (In fact, this procedure is equivalent to replacing $p_{n,m,2}$ by its steady-state value \cite{Holland96}.) Mathematically, the evolution equation for $P_{n,m}$ due to atomic injection must not depend on either of the parameters $\mu_\mathrm{in}$, $\mu_\mathrm{out}$ or, equivalently,
\begin{equation}
  \frac{d}{dt}P_{n,m}\big|_\mathrm{pump}=0.
\end{equation}
In total, the evolution equation for the atom--photon statistics $P_{n,m}$ obtained by adding Eqs.~\eqref{pnm0} and \eqref{pnm1} reads
\begin{align}
  \label{pnm}
  \frac{d}{dt}P_{n,m}&=\eta\sqrt{n m}\,V_{n,m}-\eta\sqrt{(n+1)(m+1)}\,V_{n+1,m+1}\nonumber\\
  &+2\kappa_a\left[(n+1)P_{n+1,m}-n\,P_{n,m}\right]\nonumber\\
  &+2\kappa_b\left[(m+1)P_{n,m+1}-m\,P_{n,m}\right].
\end{align}

We now want to derive a rate equation for $P_{n,m}$ in the limit where the non-diagonal elements $V_{n,m}$ can be eliminated. Whereas the relaxation time of $P_{n,m}$ is on the order of $\kappa_{a}^{-1}$, $\kappa_{b}^{-1}$, Eqs.~\eqref{pnms} evolve on a timescale which is influenced by the atomic pumping dynamics, too. When $\mu_\mathrm{in},\mu_\mathrm{out}\gg\kappa_{a,b}$, the source mode reaches its stationary state much faster than the laser modes after the creation of an atom--photon pair. The time scales are then significantly separated and one can adiabatically eliminate the fast evolving terms.

The procedure is as follows. First, neglecting all terms proportional to $\kappa_a$ and $\kappa_b$, the steady-state solution of Eqs.~\eqref{pnms} is obtained by equating the time derivatives with zero. It is then with help of Eq.~\eqref{jointprobdef} expressed in terms of the slowly-varying probabilities $P_{n,m}$ which yields
\begin{gather}
  \label{etavnm}
  \eta\sqrt{n m}\,V_{n,m}=A_{n,m}P_{n-1,m-1}-B_{n,m}P_{n,m}.
                                %\eta\sqrt{n m}\,V_{n,m}=\frac{n m R}{1+n m R}\left(\mu P_{n-1,m-1}-\mu_\mathrm{out} P_{n,m}\right),\\
                                %R=\frac{2\eta^2}{(\mu+\mu_\mathrm{out})^2}.
\end{gather}
In this equation we have defined the emission and absorption rates respectively as
\begin{subequations}
  \label{emabsrate}
  \begin{gather}
    A_{n,m}=\mu_\mathrm{in}\frac{n m R}{1+n m R},\\
    B_{n,m}=\mu_\mathrm{out}\frac{n m R}{1+n m R},
  \end{gather}
\end{subequations}
and $R=2\eta^2/(\mu_\mathrm{in}+\mu_\mathrm{out})^2$. Finally, inserting Eqs.~\eqref{etavnm} and \eqref{emabsrate} into Eq.~\eqref{pnm} we get for the rate equation
\begin{align}
  \label{rate}
  \frac{d}{dt}P_{n,m}&=B_{n+1,m+1} P_{n+1,m+1}-A_{n+1,m+1} P_{n,m}\nonumber\\
  &-B_{n,m} P_{n,m}+A_{n,m} P_{n-1,m-1}\nonumber\\
  &+2\kappa_a\left[(n+1)P_{n+1,m}-n\,P_{n,m}\right]\nonumber\\
  &+2\kappa_b\left[(m+1)P_{n,m+1}-m\,P_{n,m}\right].
                                %\dot{P}_{n,m}&=\frac{(n+1)(m+1)R}{1+(n+1)(m+1)R}\left(\mu_\mathrm{out} P_{n+1,m+1}-\mu P_{n,m}\right)\nonumber\\
                                %&-\frac{n m R}{1+n m R}\left(\mu_\mathrm{out} P_{n,m}-\mu P_{n-1,m-1}\right)\nonumber\\
                                %&+2\kappa_a\left[(n+1)P_{n+1,m}-n\,P_{n,m}\right]\nonumber\\
                                %&+2\kappa_b\left[(m+1)P_{n,m+1}-m\,P_{n,m}\right].
\end{align}
The interpretation of the individual terms is rather simple. While the last two lines originate respectively from photon loss and atomic out-coupling, the first two lines describe absorption (terms $\sim B_{n,m}$) and creation (terms $\sim A_{n,m}$) of laser atom--photon pairs. This equation is in fact very analogous to the rate equation found for a common single-mode laser showing gain and saturation.

As a central result, however, we find here joint atom--photon gain and saturation in the emission and absorption rates [Eqs.~\eqref{emabsrate}]. The origin for this twin Bose-enhancement which has already been anticipated \cite{Salzburger06,Salzburger07} is cavity-assisted cooling of atoms as realized by Eq.~\eqref{pair}. The simultaneous creation of ground-level-atom--photon pairs is further responsible for the symmetry between $n$ and $m$ in Eq.~\eqref{rate}. For equal decay rates $\kappa_a=\kappa_b$, we have perfect symmetry $P_{n,m}=P_{m,n}$ and both lasers feature identical properties.

\begin{figure}
  \includegraphics[width=8cm]{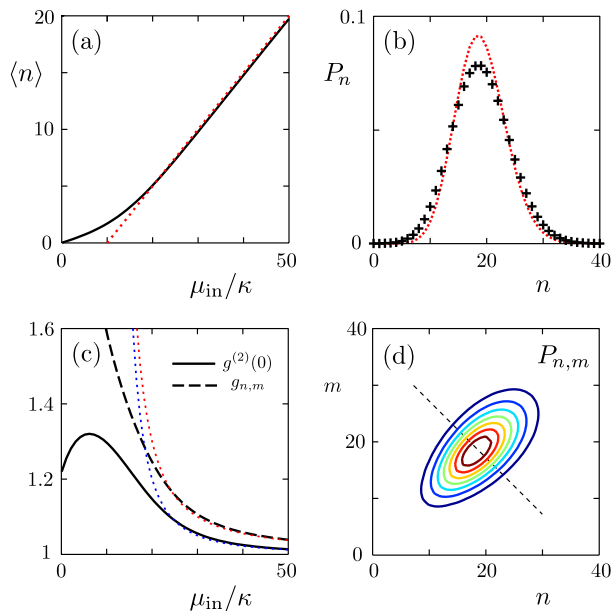}
  \caption{(a) Photon and atom mean occupation numbers $\langle n\rangle=\langle m\rangle$ in the steady state as functions of the atomic pumping rate $\mu_\mathrm{in}$. They coincide since the decay rates equal and $\kappa_a=\kappa_b=\eta/20$. The solid line and the (red) dotted line represent, respectively, the numerical solution of the rate equation \eqref{rate} and the approximate analytical solution from Eqs.\ \eqref{solanalyt} showing good agreement sufficiently far above threshold.\\
    (b) Single-mode number distribution $P_n$ (crosses) for the same parameters as in (a) for $\mu_\mathrm{in}=50\kappa$ and, for comparison, a Poisson distribution with equal mean (dotted line).\\
    (c) Single-mode correlation function $g^{(2)}(0)$ (solid line) and inter-mode correlation function $g^{n,m}$ (dashed line) as functions of the pumping rate $\mu_\mathrm{in}$ for the same parameters as in (a). The dotted lines represent the corresponding analytical results.\\
    (d) Joint excitation probability $P_{n,m}$ for the same parameters as in (b). The quenching along the broken line is significant for enlarged inter-mode correlations.}
  \label{fig1}
\end{figure}

In the following we are interested in the stationary solution of the system which can be obtained by taking the time derivative in Eq.~\eqref{rate} to zero. Before solving the resulting set of equations, let us first try to obtain analytically a solution sufficiently far above threshold where we consider the probabilities $P_{n,m}$ different from zero only for large $n$ and $m$. We are then allowed to replace $A_{n,m}$ and $B_{n,m}$ in Eq.~\eqref{rate} respectively by $\mu_\mathrm{in}$ and $\mu_\mathrm{out}$ and get the following dynamical equations for the moments $\langle n^k m^l\rangle=\sum_{n,m}n^k m^l P_{n,m}$,
\begin{subequations}
  \label{dyn}
  \begin{align}
    \dot{\langle n\rangle}=&-2\kappa_a\langle n\rangle+\mu_\mathrm{in}-\mu_\mathrm{out},\\
    \dot{\langle m\rangle}=&-2\kappa_b\langle m\rangle+\mu_\mathrm{in}-\mu_\mathrm{out},\\
    \dot{\langle n^2\rangle}=&-4\kappa_a\langle n^2\rangle+2\left(\mu_\mathrm{in}-\mu_\mathrm{out}+\kappa_a\right)\langle n\rangle\nonumber\\
      &+\mu_\mathrm{in}+\mu_\mathrm{out},\\
    \dot{\langle m^2\rangle}=&-4\kappa_b\langle m^2\rangle+2\left(\mu_\mathrm{in}-\mu_\mathrm{out}+\kappa_b\right)\langle m\rangle\nonumber\\
      &+\mu_\mathrm{in}+\mu_\mathrm{out},\\
    \dot{\langle n m\rangle}=&-2\left(\kappa_a+\kappa_b\right)\langle n m\rangle\nonumber\\
      &+\left(\mu_\mathrm{in}-\mu_\mathrm{out}\right)\left(\langle n\rangle+\langle m\rangle\right)+\mu_\mathrm{in}+\mu_\mathrm{out}.
  \end{align}
\end{subequations}
The threshold condition is here $\mu_\mathrm{in}>\mu_\mathrm{out}$. Note that Eqs.~\eqref{dyn} are only valid when $\mu_\mathrm{in}-\mu_\mathrm{out}\gg2\min\left({\kappa_{a},\kappa_b}\right)$ and should then reproduce the semiclassical results. In the stationary state, the mean photon and atom numbers are
\begin{subequations}
  \label{solanalyt}
  \begin{gather}
    \langle n\rangle=\frac{\mu_\mathrm{in}-\mu_\mathrm{out}}{2\kappa_a},\\
    \langle m\rangle=\frac{\mu_\mathrm{in}-\mu_\mathrm{out}}{2\kappa_b}.
  \end{gather}
\end{subequations}
As expected they coincide when $\kappa_a=\kappa_b$. In the remaining of this section we will discuss this situation. We will therefore write $\kappa$ instead of $\kappa_a$ and $\kappa_b$ and refer to both modes simply by $n$, too.

Figure \ref{fig1}(a) shows the mean occupation number $\langle n\rangle$ for $\kappa=0.072R=0.1\mu_\mathrm{out}$ (dotted line) as a function of the pumping strength $\mu_\mathrm{in}$ together with the numerical result obtained directly from Eq.\ \eqref{rate} (solid line). Sufficiently far above threshold we get perfect agreement. More notably, however, both modes are in a state very close to a coherent state. This is illustrated in Fig.\ \ref{fig1}(b) where the single-mode statistics $P_n=\sum_m P_{n,m}$ is plotted ($+$) compared to a corresponding Poisson distribution with equal mean value (dotted line) for $\mu_\mathrm{in}/\kappa=50$.

It is interesting to ask whether or not our system shows nonclassical features such as photon anti-bunching or noise reduction. As a measure for the number statistics we compute the second-order coherence function at $\tau=0$,
\begin{equation}
  \label{g2}
  g^{(2)}(0)=1+\frac{\langle\Delta n^2\rangle-\langle n\rangle}{\langle n\rangle^2},
\end{equation}
where $\langle\Delta n^2\rangle$ denotes the photon-number dispersion. The steady-state solution of Eqs.~\eqref{dyn} yields
\begin{equation}
  \label{corrCl}
  g^{(2)}(0)=1+\frac{\mu_\mathrm{out}}{2\kappa}\frac{1}{\langle n\rangle^2},
\end{equation}
which is represented by the (blue) dotted line in Fig.\ \ref{fig1}(c). Again, we have plotted the numerical result from the rate equation \eqref{rate} for comparison (solid line) and find agreement for large $\mu_\mathrm{in}$. There the correlation function approaches unity, the value for a coherent state. However, $g^{(2)}(0)>1$ even far above threshold, which excludes photon anti-bunching being reached.

Nevertheless, there are nonclassical effects apparent. Let us first have a look at the inter-mode correlation function
\begin{equation}
  g^{n,m}=\frac{\langle n m\rangle}{\langle n\rangle\langle m\rangle}
\end{equation}
which we can derive from the steady-state solution of Eq.~\eqref{rate}. We have depicted the result as the dashed line in Fig.\ \ref{fig1}(c). It is well known \cite{Zubairy82}, that the inter-mode correlations exceed the correlations of photons of the same beam in two-photon lasers. Mathematically, this implies violation of a corresponding Cauchy-Schwartz inequality which reads in our situation (due to the symmetry)
\begin{equation}
  g^{n,m}\leq g^{(2)}(0).
\end{equation}
Note, that here we recover the same feature:\ $g^{n,m}$ lies always above $g^{(2)}(0)$ in Fig.\ \ref{fig1}(c).

By use of the steady-state solution of Eqs.~\eqref{dyn} we can estimate that
\begin{equation}
  \label{gnm}
  g^{n,m}=g^{(2)}(0)+\frac{\mu_\mathrm{out}}{2\kappa}\frac{1}{\langle n\rangle}>g^{(2)}(0),
%  g^{n,m}=1+\frac{\mu+\mu_\mathrm{out}}{4\kappa}\frac{1}{\langle n\rangle\langle m\rangle}=g^{(2)}(0)+\frac{\mu_\mathrm{out}}{2\kappa}\frac{1}{\langle n\rangle}
\end{equation}
which indeed proves this violation, at least sufficiently far above threshold. The (red) dotted line in Fig.\ \ref{fig1}(c) corresponds to the solution of $g^{n,m}$ as derived in Eq.~\eqref{gnm}.

Another interesting feature in twin-photon lasers is noise reduction of one quadrature of the field below the classical level \cite{Milburn81}. Perfect squeezing in the stationary output-beam intensity difference was found \cite{Maia92} for which simultaneous creation of the twin photons was identified to be the origin.

Let us see whether we can find here any similarities. Figure \ref{fig1}(d) gives a contour plot of the joint probability distribution $P_{n,m}$ in the steady state indicating the afore mentioned strong inter-mode correlations. Due to the elliptic shapes of the curves we anticipate a reduced width of the intensity difference, i.e., along the broken line. From the rate equation \eqref{rate} we deduce that
\begin{subequations}
  \begin{align}
    \frac{d}{dt}\langle n-m\rangle&=-2\kappa\langle n-m\rangle\\
    \frac{d}{dt}\left\langle(n-m)^2\right\rangle&=-2\kappa\left[\left\langle(n-m)^2\right\rangle-\langle n+m\rangle\right]
  \end{align}
\end{subequations}
such that the width in the intensity difference is given by
\begin{equation}
  \left\langle(n-m)^2\right\rangle-\langle n-m\rangle^2=\frac{\langle n+m\rangle}{2},
\end{equation}
both above and below the oscillation threshold. On the other hand, classical theory predicts the inequality
\begin{equation}
  \left\langle(n-m)^2\right\rangle-\langle n-m\rangle^2\geq\langle n+m\rangle
\end{equation}
and we thus have noise reduction of $50\%$ below the classical level.  

\section{Atomic pumping by tunneling}
\label{sec:tunnel}
Based on atomic tunneling between lattice sites, we will present in what follows a model for atomic injection which provides a proper and consistent description of pumping of the system. We allow here only atoms to move within the excited band since, in general, the rates for atomic tunneling involving the lower band are usually much smaller. Let $c^\dagger$ denote the creation operator for upper-band atoms within the cavity and the operator $\Gamma$ represent the available lattice sites outside the cavity which are coupled to the inside. A corresponding tunneling Hamiltonian is
\begin{equation}
  H_\mathrm{t}=i c^\dagger\Gamma -i c\Gamma^\dagger,
  \label{tunnel}
\end{equation}
where $\Gamma$, besides reducing the number of atoms outside by one, also includes the interaction strength. Let us emphasize here that Eq.~\eqref{tunnel} can in principle also be viewed as an interaction changing the atomic state from an untrapped state to a trapped state. Atoms in the untrapped state would in this situation populate the mode represented by $\Gamma$ and be transferred to the source mode $c$ by a proper interaction. This interaction could, e.g., be a Raman transition applied on untrapped atoms which are already present in the cavity. However, to get a suitable description of pumping, we will include $H_\mathrm{t}$ into the master equation [Eq.~\eqref{rho}] and then define a system density operator by taking the trace over the levels represented by $\Gamma$.

First, we have to extend our three-mode system Hilbert space, which we may denote as $\mathcal{H}_\mathrm{sys}=\mathrm{span}\{|n\rangle\}_{n=1\ldots\infty}$, by the space $\mathcal{H}_\mathrm{out}$ spanned by the basis $\{|J\rangle\}_{J=1\ldots\infty}$, where $|J\rangle$ denotes a superposition state containing $J$ atoms. A general state can be written as $|n,J\rangle=|n\rangle\otimes|J\rangle$, where $|n\rangle\in\mathcal{H}_\mathrm{sys}$ and $|J\rangle\in\mathcal{H}_\mathrm{out}$. For the total density operator we use the notation
\begin{equation}
  \rho_\mathrm{tot}=\sum_{nmJK}\lambda_{nmJK}|n,J\rangle\langle m,K|
\end{equation}
and define our system density operator as the trace
\begin{equation}
  \rho=\sum_I\langle I|\rho_\mathrm{tot}|I\rangle=\sum_{nm}\lambda_{nm}|n\rangle\langle m|
\end{equation}
with coefficients given by the sum over the diagonal elements, $\lambda_{nm}=\sum_J\lambda_{nmJJ}$. After incorporating the tunneling Hamiltonian $H_\mathrm{t}$ into the master equation and taking the trace over $\mathcal{H}_\mathrm{out}$, the pumping term reads
\begin{gather}
  \left[c^\dagger,\sigma\right]-\left[c,\sigma^\dagger\right]
  \label{tf}
\end{gather}
where $\sigma=\sum\lambda_{nmJK}\langle K|\Gamma|J\rangle|n\rangle\langle m|.$

In the remainder of the paper we will consider the scenario where the number of outside modes is rather large and, further, these modes are in a state which is marginally affected by atom transfer to or from the intracavity system. (For example, this holds true for a coherent state or for a state with many sufficiently occupied levels.) It is then justified to approximate $\sigma\sim\mu\rho$, where $\mu$ indicates the effective pumping strength, so that from Eq.~\eqref{tf} we can read off an effective pump Hamiltonian of the form
\begin{equation}
  H_\mathrm{p}=i\mu\left(c^\dagger-c\right).
  \label{pump}
\end{equation}
Note that without loss of generality we have assumed here $\mu$ to be real thus having zero phase. It is of course an artifact of the model that the atomic field exhibits here a phase. However, Eq.~\eqref{tunnel} conserves the atom number such that the phase has no observable consequences. $H_\mathrm{p}$ thus provides a self-consistent description of continuous atomic occupation of the source mode originating from tunneling.

In addition to pair generation [Eq.~\eqref{pair}] and atom injection [Eq.~\eqref{pump}] let us also include atomic heating into the dynamics. Heating should regard both the atomic laser mode---atoms are heated into the source mode at rate $\kappa_{bc}$ there---and the source mode, which we model simply by a depletion channel (rate $\kappa_c$). The full master equation reads then \cite{qn}
\begin{align}
  \dot{\rho}=&-i[H_\mathrm{int}+H_\mathrm{p},\rho]\nonumber\\
  &+\kappa_a\left(\left[a,\rho\,a^\dagger\right]+\left[a\,\rho,a^\dagger\right]\right)\nonumber\\
  &+\kappa_b\left(\left[b,\rho\,b^\dagger\right]+\left[b\,\rho,b^\dagger\right]\right)\nonumber\\
  &+\kappa_{bc}\left(\left[c^\dagger b,\rho\,b^\dagger c\right]+ \left[c^\dagger b\,\rho,b^\dagger c\right]\right)\nonumber\\
  &+\kappa_c\left(\left[c,\rho\,c^\dagger\right]+\left[c\,\rho,c^\dagger\right]\right).
  \label{me}
\end{align}

We proceed by converting the master equation \eqref{me} into a Fokker-Planck equation (FPE) for the generalized $P$-representation of the density operator. The generalized $P$-representation is defined as
\begin{equation}
  \rho=\int\mathrm{d}\boldsymbol{\alpha}\,\mathrm{d}\boldsymbol{\alpha}^\dagger\,P(\boldsymbol{\alpha},\boldsymbol{\alpha}^\dagger)\frac{|\boldsymbol{\alpha}\rangle\langle{\boldsymbol{\alpha}^\dagger}^*|}{\langle{\boldsymbol{\alpha}^\dagger}^*|\boldsymbol{\alpha}\rangle},
\end{equation}
where $\boldsymbol{\alpha}=(\alpha,\beta,\gamma)$ and $\boldsymbol{\alpha}^\dagger=(\alpha^\dagger,\beta^\dagger,\gamma^\dagger)$ are independent complex variables. Applying standard operator identities \cite{qn}, the FPE for the $P$-representation reads
\begin{gather}
  \frac{\partial}{\partial t}P(\boldsymbol{\alpha},\boldsymbol{\alpha}^\dagger)=-\frac{\partial}{\partial\alpha}\left(-\kappa_a\alpha+\eta\,\beta^\dagger\gamma\right)P(\boldsymbol{\alpha},\boldsymbol{\alpha}^\dagger)\nonumber\\
  -\frac{\partial}{\partial\beta}\left[-\kappa_b\beta-\kappa_{bc}\beta\left(\gamma^\dagger\gamma+1\right)+\eta\,\alpha^\dagger\gamma\right]P(\boldsymbol{\alpha},\boldsymbol{\alpha}^\dagger)\nonumber\\
  -\frac{\partial}{\partial\gamma}\left(-\kappa_c\gamma+\kappa_{bc}\beta^\dagger\beta\gamma-\eta\,\alpha\beta+\mu\right)P(\boldsymbol{\alpha},\boldsymbol{\alpha}^\dagger)\nonumber\\
  +\frac{\partial^2}{\partial\alpha\partial\beta}\eta\,\gamma\,P(\boldsymbol{\alpha},\boldsymbol{\alpha}^\dagger)+\frac{\partial^2}{\partial\beta\partial\gamma}\left(-\kappa_{bc}\beta\,\gamma\right)P(\boldsymbol{\alpha},\boldsymbol{\alpha}^\dagger)\nonumber\\
  +\frac{\partial^2}{\partial\gamma\partial\gamma^\dagger}\kappa_{bc}\beta^\dagger\beta P(\boldsymbol{\alpha},\boldsymbol{\alpha}^\dagger)+\mathrm{conj.}
  \label{fpe}
\end{gather}
where the conjugate terms are given by the formal replacement $\boldsymbol{\alpha}\leftrightarrow\boldsymbol{\alpha}^\dagger$. The use of the $P$-representation enables us to derive normally ordered expectation values of quantum operators simply as the corresponding moments of $P(\boldsymbol{\alpha},\boldsymbol{\alpha}^\dagger)$. However, we will not present here a solution to the FPE \eqref{fpe} but use an equivalent set of stochastic differential equations (SDE) for the variables $\boldsymbol{\alpha}$ and $\boldsymbol{\alpha}^\dagger$. Therefore, we can identify quantum expectation values of any polynomial $p$ composed of the operators $\boldsymbol{a}=(a,b,c)$ with the mean of our stochastic variables as
\begin{equation}
  \left\langle:\!p\left(\boldsymbol{a},\boldsymbol{a}^\dagger\right)\!\!:\right\rangle_\mathrm{qu}=\left\langle p\left(\boldsymbol{\alpha},\boldsymbol{\alpha}^\dagger\right)\right\rangle_\mathrm{st},
\end{equation}
where the double dots indicate normal ordering. As an indication for normal order we will later express the quantities of interest in terms of the intensities $I_{\boldsymbol{\alpha}}=\boldsymbol{\alpha}^\dagger\boldsymbol{\alpha}$. (Note that the bold variables are used here as a formal placeholder for one of the variables $\alpha$, $\beta$, and $\gamma$ at a time.) This causes normally ordered expectation values to be calculated simply as
\begin{equation}
  \label{qu2st}
  \langle:\!\hat{n}_{\boldsymbol{a}}^m\!:\rangle_\mathrm{qu}=\langle I_{\boldsymbol{\alpha}}^m\rangle_\mathrm{st},
\end{equation}
where $\hat{n}_{\boldsymbol{a}}$ denotes the corresponding quantum number operator.

The SDE have the same drift and diffusion terms as the FPE \eqref{fpe} and read
\begin{subequations}
  \label{sde}
  \begin{gather}
    \dot{\alpha}=-\kappa_a\alpha+\eta\beta^\dagger\gamma+\xi_\alpha,\\
    \dot{\beta}=-\kappa_b\beta-\kappa_{bc}\beta\left(\gamma^\dagger\gamma+1\right)+\eta\alpha^\dagger\gamma+\xi_\beta,\\
    \dot{\gamma}=-\left(\kappa_c-\kappa_{bc}\beta^\dagger\beta\right)\gamma - \eta\,\alpha\beta+\mu+\xi_\gamma,
  \end{gather}
\end{subequations}
where the $\xi_{\boldsymbol{\alpha}}$ are $\delta$-correlated noise terms with zero mean and the following non-zero diffusion coefficients,
\begin{subequations}
  \label{diffcoeffs}
  \begin{gather}
    \mathcal{D}_{\alpha\beta}=\eta\langle\gamma\rangle,\\
    \mathcal{D}_{\alpha^\dagger\beta^\dagger}=\eta\langle\gamma^\dagger\rangle,\\
    \mathcal{D}_{\beta\gamma}=-\kappa_{bc}\langle\beta\gamma\rangle,\\
    \mathcal{D}_{\beta^\dagger\gamma^\dagger}=-\kappa_{bc}\langle\beta^\dagger\gamma^\dagger\rangle,\\
    \mathcal{D}_{\gamma^\dagger\gamma}=\kappa_{bc}\langle\beta^\dagger\beta\rangle.
  \end{gather}
\end{subequations}

Before we proceed, let us first investigate the semiclassical behavior of the system. It is gained by ignoring the stochastic nature of the system variables and equations, i.e., by taking $\boldsymbol{\alpha}^\dagger\equiv\boldsymbol{\alpha}^*$ and $\xi_{\boldsymbol{\alpha}}\equiv 0$ in Eqs.~\eqref{sde}.

The steady-state solution $\langle\boldsymbol{\alpha}\rangle=\boldsymbol{\alpha}_0$ can be obtained by equating the drift terms in Eqs.~\eqref{sde} with zero. As mentioned, we chose for convenience the pumping strength $\mu$ to be real so that the semiclassical solution $\boldsymbol{\alpha}_0$ has no phase. The result shows a threshold appearing at
\begin{equation}
  \mu_\mathrm{th}=\kappa_c\sqrt{\frac{\kappa_a(\kappa_b+\kappa_{bc})}{\eta^2-\kappa_a\kappa_{bc}}} 
\end{equation}
for $\eta^2>\kappa_a\kappa_{bc}$, below which the mean field amplitudes are $\alpha_0=\beta_0=0$ and $\gamma_0=\mu/\kappa_c$, while above we have
\begin{subequations}
  \label{semi}
  \begin{gather}
    \alpha_0=\frac{\eta\sqrt{\kappa_c(\kappa_b+\kappa_{bc})}}{\eta^2-\kappa_a\kappa_{bc}}\sqrt{\epsilon-1},\\
    \beta_0=\sqrt{\frac{\kappa_a\kappa_c}{\eta^2-\kappa_a\kappa_{bc}}}\sqrt{\epsilon-1},\\
    \gamma_0=\frac{\mu_\mathrm{th}}{\kappa_c},
  \end{gather}
\end{subequations}
where $\epsilon=\mu/\mu_\mathrm{th}$ is the pumping parameter. The mean occupation numbers are then $n_{\boldsymbol{a}}=\boldsymbol{\alpha}_0^2$, respectively.

The semiclassical solution $\langle\boldsymbol{\alpha}\rangle$ may provide an approximation to the solution of the FPE \eqref{fpe} for small fluctuations. As an example, for lasers this is fulfilled far above threshold \cite{Davidovich96}. Nevertheless, it does not account for correlations of any kind and, in general, high-order expectation values of $\boldsymbol{\alpha}$ disagree with the semiclassical counterpart. One is therefore obliged to include the fluctuations into the dynamics in order to calculate for instance the number statistics or output spectra of the laser.

In what follows, we will again take the noise terms $\xi_{\boldsymbol{\alpha}}$ into account and derive equations for the dynamics of the fluctuations around the steady-state solutions $\boldsymbol{\alpha}_0$. These quantities turn out to be directly related to the noise spectra from which we can calculate higher-order moments in an appropriate way.

\section{Dynamics of fluctuations}
\label{output}
To account properly for the fluctuations, we now express the dynamic variables as a sum of the steady-state values plus small fluctuations,
\begin{equation}
  \label{alpha}
  \boldsymbol{\alpha}=\boldsymbol{\alpha}_0+\delta\boldsymbol{\alpha},
\end{equation}
(and analogously for $\boldsymbol{\alpha}^\dagger$) and linearize the equations of motion around their steady-state values. Note that it is important to be in a parameter regime where the fluctuations of the dynamical variables are much smaller than their steady-state values, in order to justify this assumption. We will later see that this criterion can be fulfilled even not very far above the threshold.

After substitution of Eq.~\eqref{alpha} into the SDE \eqref{sde}, the linearized equations for the fluctuations $\delta\boldsymbol{\alpha}$ and $\delta\boldsymbol{\alpha}^\dagger$ are
\begin{equation}
  \label{flucts}
  \frac{\mathrm{d}}{\mathrm{d}t}\left[\begin{array}{c}\delta\boldsymbol{\alpha}\\\delta\boldsymbol{\alpha}^\dagger\end{array}\right]
  =\mathbf{M}\left[\begin{array}{c}\delta\boldsymbol{\alpha}\\\delta\boldsymbol{\alpha}^\dagger\end{array}\right]+\left[\begin{array}{c}\xi_{\boldsymbol{\alpha}}\\\xi_{\boldsymbol{\alpha}^\dagger}\end{array}\right],
\end{equation}
where the matrix $\mathbf{M}$ is composed of four blocks,
\begin{equation}
  \mathbf{M}=\left[\begin{array}{c|c}A&B\\\hline B&A\end{array}\right],
\end{equation}
with
\begin{equation}
    A=\left[\begin{array}{ccc}-\kappa_a&0&\eta\beta_0\\0&-\kappa_b-\kappa_{bc}(\gamma_0^2+1)&\eta\alpha_0-\kappa_{bc}\beta_0\gamma_0\\-\eta\beta_0&-\eta\alpha_0+\kappa_{bc}\beta_0\gamma_0&-\kappa_c+\kappa_{bc}\beta_0^2\end{array}\right]
\end{equation}
and
\begin{equation}
  B=\left[\begin{array}{ccc}0&\eta\gamma_0&0\\\eta\gamma_0&0&-\kappa_{bc}\beta_0\gamma_0\\0&\kappa_{bc}\beta_0\gamma_0&0\end{array}\right].
\end{equation}
The solution of Eq.\ \eqref{flucts} is best found by a Fourier transform method. Let us define the variables $\boldsymbol{v}=(\delta\boldsymbol{\alpha},\delta\boldsymbol{\alpha}^\dagger,\xi_{\boldsymbol{\alpha}},\xi_{\boldsymbol{\alpha}^\dagger})$ in frequency space as
\begin{gather}
  \boldsymbol{v}(\omega)=\frac{1}{\sqrt{2\pi}}\int \mathrm{d}\tau\,\mathrm{e}^{i\omega\tau}\boldsymbol{v}(\tau)
  \label{defnewvars}
\end{gather}
and note that we will for convenience use the same symbol both in time space and in frequency space while, though, the argument should make the distinction obvious. It follows then from this definition that the noise terms $\xi_{\boldsymbol{\alpha}}(\omega)$ obey the correlations
\begin{equation}
  \label{omegacorrs}
  \langle\xi_{\boldsymbol{\alpha}}(\omega)\xi_{\boldsymbol{\beta}}(\omega')\rangle=\mathcal{D}_{\boldsymbol{\alpha\beta}}\,\delta(\omega+\omega')
\end{equation}
with the same diffusion coefficients as the untransformed variables given in Eqs.~\eqref{diffcoeffs}. Applying the Fourier transform on Eq.~\eqref{flucts} yields in the steady state
\begin{equation}
  \label{flucts2}
  \left[\begin{array}{c}\delta\boldsymbol{\alpha}(\omega)\\\delta\boldsymbol{\alpha}^\dagger(\omega)\end{array}\right]
  =\left(-i\omega-\mathbf{M}\right)^{-1}\left[\begin{array}{c}\xi_{\boldsymbol{\alpha}}(\omega)\\\xi_{\boldsymbol{\alpha}^\dagger(\omega)}\end{array}\right],
\end{equation}
which is an algebraic problem that can be solved exactly. The solution in time space is then---of course---governed by the inverse Fourier transform.

Using Eqs.~\eqref{omegacorrs} and \eqref{flucts2}, we define the spectral functions $A_{\boldsymbol{\alpha\beta}}(\omega)$ and $Q_{\boldsymbol{\alpha\beta}}(\omega)$ written in terms of the quadrature component $\delta x_{\boldsymbol{\alpha}}=\delta\boldsymbol{\alpha}+\delta\boldsymbol{\alpha}^\dagger$ as
\begin{subequations}
  \label{aandq}
  \begin{gather}
    \langle\delta\boldsymbol{\alpha}^\dagger(\omega)\delta\boldsymbol{\beta}(\omega')\rangle=A_{\boldsymbol{\alpha\beta}}(\omega)\delta(\omega+\omega'),\\
    \langle\delta x_{\boldsymbol{\alpha}}(\omega)\delta x_{\boldsymbol{\beta}}(\omega')\rangle=Q_{\boldsymbol{\alpha\beta}}(\omega)\delta(\omega+\omega').
  \end{gather}
\end{subequations}
They are related to the spectra of, respectively, the amplitude and quadrature fluctuations by just adding the corresponding shot-noise contributions. We will come back to this point later when we discuss number statistics and squeezing.

According to Eq.~\eqref{alpha} the mean populations $n_{\boldsymbol{a}}=\langle I_{\boldsymbol{\alpha}}\rangle$ are given by
\begin{equation}
  \label{correction2semi}
  \langle \boldsymbol{\alpha}^\dagger\boldsymbol{\alpha}\rangle=\boldsymbol{\alpha}_0^2+\langle\delta\boldsymbol{\alpha}^\dagger\delta\boldsymbol{\alpha}\rangle,
%  \langle\boldsymbol{\alpha}^\dagger\boldsymbol{\alpha}\rangle=\boldsymbol{\alpha}_0^2\left(1+\langle\delta\boldsymbol{\alpha}^\dagger\delta\boldsymbol{\alpha}\rangle\right),
\end{equation}
where we have used the fact that $\langle\delta\boldsymbol{\alpha}\rangle=\langle\delta\boldsymbol{\alpha}^\dagger\rangle=0$. The correction to the semiclassical solution is obtained by the integral over the amplitude fluctuation spectrum,
\begin{equation}
  \langle\delta\boldsymbol{\alpha}^\dagger\delta\boldsymbol{\alpha}\rangle=\frac{1}{2\pi}\int\mathrm{d}\omega\,A_{\boldsymbol{\alpha\alpha}}(\omega).
\end{equation}
The result for the photon laser intensity $\langle I_{\alpha}\rangle$ is shown as the dash-dotted (blue) line in Fig.\ \ref{fig2}(a) as a function of the pumping parameter $\epsilon$. The remaining parameters are $(\eta,\kappa_{bc},\kappa_c)=(4,0,2)\kappa_a$. Therefore the solutions for both modes $a$ and $b$ coincide for equal decay rates $\kappa_a=\kappa_b$. The mean number follows closely the linear semiclassical result which we have indicated by the crosses.

However, care must be taken for $\mu\approx\mu_\mathrm{th}$ since the fluctuations around the steady-state may be there too large to justify the linearization assumption. For comparison we have also plotted the quantity $\langle\delta\alpha^\dagger\delta\alpha\rangle$ as the dotted (red) line which turns out to be much less than one even not so far above the threshold.

\subsection{Number statistics for equal decay rates}
Laser light is distinguished by having small intensity fluctuations in the sense that the standard deviation of the intensity is small compared to its mean. The fluctuations are usually investigated by means of the variance of the photon number. It is well known that the light emitted from a laser operating far above the oscillation threshold exhibits the statistics of independent particles, which corresponds to a Poisson distribution for the number of photons in the laser mode.

Let us write down the photon-number variance in normal order which reads
\begin{equation}
  \label{dn2}
  \langle\Delta \hat{n}_{\boldsymbol{a}}^2\rangle=\langle\hat{n}_{\boldsymbol{a}}\rangle+\langle:\!\Delta\hat{n}_{\boldsymbol{a}}^2\!:\rangle.
\end{equation}
The first term on the right-hand side of this equation originates from the commutator of the bosonic operators $\boldsymbol{a}$ and $\boldsymbol{a}^\dagger$ and reflects the quantum character of the field. In fact, it corresponds to the contribution to the number variance of an ensemble of independent particles. Commonly called the shot noise, it is in principle inherent to any state of the electromagnetic field due to the discreteness of the photon-number distribution. For Poissonian number statistics it constitutes the sole contribution to the number variance, and field states for which the photon-number variance is less than the shot-noise term are necessarily nonclassical \cite{Davidovich96}.

As a consequence of using the $P$-function representation, which is mathematically manifested in equality \eqref{qu2st}, the second term on the right-hand side of Eq.~\eqref{dn2} is equal to the c-number intensity dispersion,
\begin{equation}
  \label{dn2di2}
  \langle:\!\Delta\hat{n}_{\boldsymbol{a}}^2\!:\rangle=\langle\Delta I_{\boldsymbol{\alpha}}^2\rangle,
\end{equation}
where
\begin{equation}
  \label{di2}
  \langle\Delta I_{\boldsymbol{\alpha}}^2\rangle=\langle I_{\boldsymbol{\alpha}}^2\rangle-\langle I_{\boldsymbol{\alpha}}\rangle^2.
\end{equation}
This term exhibits the characteristic of the $P$-function playing the role of a quasi-classical distribution function. We can hence associate the deviation from the Poisson distribution with the intensity dispersion so that negative and positive $\langle\Delta I_{\boldsymbol{\alpha}}^2\rangle$ indicate sub- and super-Poissonian statistics, respectively.

%The second term in Eq.~\eqref{dn2}, which can be associated with the deviation from the Poisson distribution, is equal to the (normally-ordered) intensity dispersion
%\begin{equation}
%  \label{di2}
%  \langle\Delta I_{\boldsymbol{\alpha}}^2\rangle=\left\langle\left(I_{\boldsymbol{\alpha}}-\langle I_{\boldsymbol{\alpha}}\rangle\right)^2\right\rangle.
%\end{equation}
Intensity-correlation experiments reveal the cumulation tendency of photons of a light beam. Usually, one is interested in the joint probability of detecting a photon respectively at times $t$ and $t+\tau$. In the stationary state, the probability is proportional to the second-order coherence function
\begin{equation}
  \label{g2neu}
  g^{(2)}(\tau)=\frac{\langle:\!\hat{n}(t)\hat{n}(t+\tau)\!:\rangle}{\langle\hat{n}\rangle^2}.
\end{equation}
Note, that one would have $g^{(2)}(\tau)=1$ for independent particles while $g^{(2)}(0)<g^{(2)}(\tau)$ is significant for photon anti-bunching.

On the other hand, there exists a relation between the degree of second-order coherence and the number statistics. Obviously, the numerator in Eq.~\eqref{g2neu} is equal to the normally-ordered expectation value of $\hat{n}^2$ if $\tau=0$. It is thus possible to re-express the coherence function for $\tau=0$ in terms of the number variance. As before we use here the intensity variance (to account for the normal order) and get
\begin{equation}
  \label{g3neu}
  g_{\boldsymbol{a}}^{(2)}(0)=1+\frac{\langle\Delta I_{\boldsymbol{\alpha}}^2\rangle}{\langle I_{\boldsymbol{\alpha}}\rangle^2}.
\end{equation}
So far no approximation has been made. The intensity dispersion in Eq.~\eqref{g3neu} involves fourth-order correlations of the field variables $\boldsymbol{\alpha}$. Nevertheless, in a regime of small fluctuations, one can compute this quantity in terms of the amplitude quadrature variance which requires only second-order correlations. Analogous to Eq.~\eqref{alpha} we therefore write for the intensity $I_{\boldsymbol{\alpha}}=\langle I_{\boldsymbol{\alpha}}\rangle+\delta I_{\boldsymbol{\alpha}}$ so that the variance in the intensity is equal to $\langle\left(\delta I_{\boldsymbol{\alpha}}\right)^2\rangle$. We further assume that the classical relation between intensity and amplitude fluctuations
\begin{equation}
  \label{di}
  \delta I_{\boldsymbol{\alpha}}\approx\boldsymbol{\alpha}_0\left(\delta\boldsymbol{\alpha}^\dagger+\delta\boldsymbol{\alpha}\right)=\boldsymbol{\alpha}_0\delta x_{\boldsymbol{\alpha}}
\end{equation}
holds true. Note that this approximation is justified as long as $\langle\delta\boldsymbol{\alpha}^\dagger\delta\boldsymbol{\alpha}\rangle^2\ll\boldsymbol{\alpha}_0^2\langle\delta x_{\boldsymbol{\alpha}}^2\rangle$. We get in this case for the intensity dispersion
\begin{equation}
  \label{di3}
  \langle\Delta I_{\boldsymbol{\alpha}}^2\rangle=\boldsymbol{\alpha}_0^2\langle\delta x_{\boldsymbol{\alpha}}^2\rangle
\end{equation}
where, written in terms of the spectral functions \eqref{aandq},
\begin{equation}
  \label{dx2}
  \langle\delta x_{\boldsymbol{\alpha}}^2\rangle=\frac{1}{2\pi}\int\mathrm{d}\omega\,Q_{\boldsymbol{\alpha\alpha}}(\omega).
\end{equation}

\begin{figure}
  \includegraphics[width=8cm]{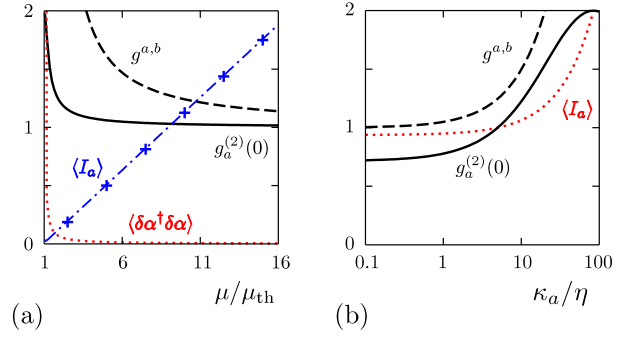}
  \caption{(a) Single-mode correlation function $g_a^{(2)}(0)$ (solid line) and inter-mode correlation function $g^{a,b}$ (dashed line) where the variable is the pumping strength $\mu$. In addition we have plotted the mean intensity $\langle I_\alpha\rangle$ (blue dash-dotted line) together with the corresponding semiclassical result (+), and the quantity $\langle\delta\alpha^\dagger\delta\alpha\rangle$ (red dotted line).\\
  (b) Dependence of the correlation functions $g_a^{(2)}(0)$ (solid line) and $g^{a,b}$ (dashed line) on the out-coupling rates $\kappa_a=\kappa_b$. The pumping strength $\mu$ is here adjusted to yield mean intensities of around one as indicated by the (red) dotted line.}
  \label{fig2}
\end{figure}

When $|\langle\delta x_{\boldsymbol{\alpha}}^2\rangle|\ll1$, the intensity dispersion can be neglected compared to its mean. According to Eqs.~\eqref{dn2} and \eqref{dn2di2}, the photon number variance is in this case $\langle\Delta\hat{n}_{\boldsymbol{a}}^2\rangle=\langle\hat{n}_{\boldsymbol{a}}\rangle$ which is typical of Poissonian number statistics. Equivalently, the second-order coherence function at $\tau=0$ equals one, which follows from Eq.~\eqref{g3neu}. The $P$-function is then very close (or equal) to a $\delta$-distribution representing the density matrix of a coherent state. This behavior is demonstrated in Fig.\ \ref{fig2}(a) sufficiently far above threshold. The solid line shows the coherence function $g^{(2)}_a(0)$. We have chosen here a set of parameters such that the solutions for modes $a$ and $b$ coincide, i.e., $\kappa_a=\kappa_b=\kappa_c/2=\eta/4$. At and slightly above threshold, the system is in a state close to a thermal state and $g^{(2)}_a(0)\lesssim2$. With increasing pumping strength the laser modes start to become populated. It is significant for the stimulated enhancement of the desired processes that the coherence function drops at the same time---quite sharply---to its asymptotic value. For the parameters chosen here it is unity which corresponds to a coherent state.

In addition we have also plotted the inter-mode correlation function, which is defined as
\begin{equation}
  g^{a,b}=\frac{\langle\alpha^\dagger\alpha\beta^\dagger\beta\rangle}{\langle\alpha^\dagger\alpha\rangle\langle\beta^\dagger\beta\rangle},
\end{equation}
as the broken line. As a manifestation of the pair creation we find large correlations between atoms and photons. Actually, the curve for $g^{a,b}$ lies always above the curve that belongs to the coherence functions, which implies that the mutual correlations between atoms and photons are much larger than the correlations between the particles of one and the same beam are.

Let us mention at this point that Eq.~\eqref{di} relates number squeezing to amplitude quadrature squeezing. Amplitude and number squeezing are not equivalent. As an example, amplitude squeezed states can have both super- and sub-Poissonian number statistics for fixed photon number, depending solely on the squeezing parameter \cite{Davidovich96}.

Equation \eqref{dx2} is proportional to the normally ordered quantum expectation value of the variance of the field amplitude quadrature $\boldsymbol{X}=\boldsymbol{a}+\boldsymbol{a}^\dagger$, namely
\begin{equation}
  \langle\delta x_{\boldsymbol{\alpha}}^2\rangle=\langle:\!\Delta\boldsymbol{X}^2\!:\rangle.
\end{equation}
Amplitude quadrature squeezing, where $\langle:\!\Delta\boldsymbol{X}^2\!:\rangle<0$, is therefore only possible when the corresponding $P$-function is not positive definite. As a consequence, the intensity dispersion \eqref{di3} is negative and the modes exhibit sub-Poissonian statistics. Mathematically, the coherence function $g^{(2)}_{\boldsymbol{a}}(0)<1$.

This situation can be achieved by reducing the decay rates as is demonstrated in Fig.\ \ref{fig2}(b). It shows the coherence function $g^{(2)}_a(0)$ (solid line) and the correlation function $g^{a,b}$ (broken line) for $\kappa_b=\kappa_a$. As already mentioned in the discussion of Fig.\ \ref{fig15}, reduction of the out-coupling rates induces stimulated enhancement of the twin gain with large atom-laser emission rate for a cavity with high quality. In principle, we find here the same results. However, here we would like to keep the mode occupations constant for a more refined comparison. Therefore the pumping strength $\mu$ is adjusted here to achieve mean intensities $\langle I_\alpha\rangle$ of the order of one by use of the semiclassical result. In the figure, $\langle I_\alpha\rangle$ is indicated by the dotted line. The other parameters are $\eta=10\kappa_c$ and $\kappa_{bc}=0$ so that the results for modes $a$ and $b$ coincide. For large $\kappa_a$, the second-order correlation function starts from two where the mode is in a thermal state. It decreases when the linewidth is reduced and eventually drops below unity. We have checked that at the same time the threshold pumping strength $\mu_\mathrm{th}$ decreases linearly with the cavity line width. A good cavity thus not only enhances the emission of ground-level atoms but also eases the requirements on the pumping strength (i.e., the atomic density) to reach degeneracy.

Note finally that the inter-mode correlation $g^{a,b}$ is again always larger than $g^{(2)}(0)$ and reaches an asymptotic value of 1.

\subsection{Output-field spectra}
So far we have focused on the internal properties of our system and were able, for instance, to derive the mean number and the statistics of the laser modes. In the scope of any experiment, however, one is usually interested in accessible quantities such as the output fields. We will therefore discuss now the properties of the output beams, particularly the fluctuation spectra of the output fields and of the intensity difference.

\begin{figure}
  \includegraphics[width=8cm]{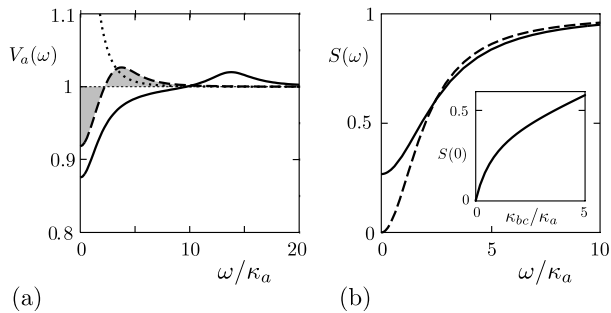}
  \caption{(a) Fluctuation spectrum $V_a(\omega)$ for different values of the pumping parameter $\epsilon$.
    The areas below the curves are directly correlated with the number statistics (see text).
    Sufficiently far above the threshold (solid line) the system exhibits amplitude noise reduction and sub-Poisson statistics.\\
    (b) Spectrum of the intensity-difference fluctuations in the output beams, $S(\omega)$, for $\kappa_{bc}=0$ (dashed line) and $\kappa_{bc}=\kappa_a$ (sold line).
    The inset shows the contribution at zero frequency $S(0)$ as a function of the heating rate $\kappa_{bc}$.
}
  \label{fig3}
\end{figure}

The result \eqref{di} which applies for instance for an laser operating far above the threshold not only connects quadrature and number squeezing. Moreover, it gives a relation between the number statistics of the internal fields and the spectra of the output fields which are accessible in experiments. The fluctuation spectrum in the field is given by the Wiener-Khinchine theorem \cite{Louisell73}. We will use here the input--output formalism developed by Collet and Gardiner \cite{Collet84} to express the output-field operators $\boldsymbol{b}_\mathrm{out}(t)$ in terms of the system variables. The spectrum of fluctuations corresponding to the quadrature $\boldsymbol{X}=\boldsymbol{b}_\mathrm{out}+\boldsymbol{b}_\mathrm{out}^\dagger$ can then be written as
\begin{equation}
  \label{spec1}
  V_{\boldsymbol{a}}(\omega)=1+2\kappa_{\boldsymbol{a}}\int\mathrm{d}\tau\,\mathrm{e}^{i\omega\tau}\langle\delta x_{\boldsymbol{\alpha}}(\tau)\delta x_{\boldsymbol{\alpha}}(0)\rangle.
\end{equation}
The first term on the right-hand side of this expression originates from the commutator of the operators $\boldsymbol{b}_\mathrm{out}$ and $\boldsymbol{b}_\mathrm{out}^\dagger$ when putting them into normal order and corresponds to the shot noise. Equation \eqref{spec1} may also be expressed in terms of the spectral functions Eqs.~\eqref{aandq} which yields
\begin{equation}
  \label{spec2}
  V_{\boldsymbol{a}}(\omega)=1+2\kappa_{\boldsymbol{a}}Q_{\boldsymbol{\alpha\alpha}}(\omega).
\end{equation}
As we have seen previously, sub-Poissonian photon statistics is a consequence of amplitude-quadrature squeezing. According to Eqs.~\eqref{dx2} and \eqref{spec2}, one must have in this case $V_{\boldsymbol{a}}(\omega)<1$, at least in some frequency range. This type of noise compression is illustrated in Fig.\ \ref{fig3}(a) showing the fluctuation spectrum $V_a(\omega)$ for different values of the pumping parameter $\epsilon$. The other parameters are the same as in Fig.\ \ref{fig2}(b) with $\kappa_a=\kappa_c/5$. Close to the threshold, the spectrum $V_a(\omega)$ lies above one as is indicated by the dotted line for which $\epsilon=1.5$. With increasing pump, noise reduction in the low-frequency range occurs until eventually the contributions of noise attenuation and accentuation cancel. This is illustrated by the gray areas in the plot for which $\epsilon=2.7$ and $g^{(2)}(0)=1$ (dashed line). Far above the threshold the statistics is sub-Poissonian, the solid line in the figure corresponds to $\epsilon=20$ with $g^{(2)}(0)=0.85$.

As a result of simultaneous atom--photon pair generation, the atom and photon numbers in the lasing modes are strongly correlated. They are, actually, exactly equal so long as either an atom or a photon is removed from the system. The variance of the occupation-number difference vanishes therefore in a system without losses.

For a steady operation including reloading and depletion of atoms we already found in Sec.~\ref{rates} a reduction of the variance in the number difference. Provided that all atoms and photons are detected, this feature should be recovered for the output intensities, as is the case for a parametric amplifier. In parametric amplification which our system is very closely related to, perfect noise reduction at zero frequency was found in the output spectrum of intensity-difference fluctuations.

Let us see what this means here. The fluctuation spectrum in the output intensity-difference is given by
\begin{equation}
  \label{S}
  S(\omega)=\int\mathrm{d}\omega\,\mathrm{e}^{i\omega\tau}\left[\langle\hat{\Delta}(\tau)\hat{\Delta}(0)\rangle-\langle\hat{\Delta}\rangle^2\right],
\end{equation}
where $\hat{\Delta}=\hat{a}^\dagger_\mathrm{out}\hat{a}_\mathrm{out}-\hat{b}^\dagger_\mathrm{out}\hat{b}_\mathrm{out}$. As before we can express the spectrum in terms of the internal variables and get, after putting the right-hand side of Eq.~\eqref{S} into normal order,
\begin{gather}
  S(\omega)=2\kappa_a\langle I_a\rangle+2\kappa_b\langle I_b\rangle%\nonumber\\
  +(2\kappa_a\alpha_0)^2 Q_{\alpha\alpha}(\omega)\nonumber\\+(2\kappa_b\beta_0)^2 Q_{\beta\beta}(\omega)
  -8\kappa_a\kappa_b\alpha_0\beta_0Q_{\alpha\beta}(\omega).
  \label{S1}
\end{gather}
Here, the first two terms are the shot noise. Figure \ref{fig3}(b) displays the normalized fluctuation spectrum \eqref{S1} divided by the shot-noise contribution. When the heating rate $\kappa_{bc}=0$ vanishes (dashed line), we find perfect noise reduction at resonance. Even for unequal output-coupling rates $\kappa_a\neq\kappa_b$, the zero-frequency contribution $S(0)$ vanishes and the system still realizes a perfectly quantum correlated pair source.

Nevertheless, atomic scattering due to collisional heating or photon emission into modes other than the cavity mode can cause deviations in the properties of the output fluxes. For an illustration we have chosen the atomic heating rate $\kappa_{bc}$ as high as the out-coupling rates $\kappa_a=\kappa_b$ for the solid line in the plot. While there is still noise reduction in the low-frequency range, the noise may be slightly increased for higher frequencies (not shown in the plot). The contribution at resonance is shown in the inset as a function of the heating rate $\kappa_{bc}$.

\section{Conclusions}
We have shown that by help of an optical cavity one can implement the final cooling stage of an atom laser via Raman sideband gain. As the corresponding anti-Stokes transitions are enhanced by stimulated emission through the photons and atoms present, the setup should allow for the realization of an atom--photon twin laser. Generation of cavity photons is always accompanied by atomic transitions to the atom laser mode, such that photon emission through the mirror acts as a diode for atom accumulation in a single state. The more photons present the more atoms are transferred to the lower state on average. Hence this device produces two output beams consisting of photons and atoms, respectively, with genuine properties.

Compared to conventional sideband cooling, the use of a cavity reduces the requirements on the atomic density (threshold for degeneracy) and avoids the problems of photon spontaneous emission and reabsorption as well as atom--atom interactions which are detrimental to the atom-laser phase \cite{Holland96}. When quantum degeneracy is reached, an atomic beam is provided by a proper out-coupling mechanism.

Using a simple model for incoherent atomic pumping we were able to derive from the master equation a rate equation for the joint atom--photon number distribution very analogous to the usual laser equations, but even more reminiscent of an optical parametric oscillator operated in the strong coupling limit \cite{Alge97}. Our model allowed to explicitly calculate absorption and creation rates of atom--photon pairs. Analogous to a laser the above-threshold population of both laser modes yields a close Poissonian number statistics in the steady state with even some anti-bunching possible for the atoms. The correlations between particles and photons are significantly larger and even show squeezing in the number difference. In analogy to a twin photon source we thus could get a heralded single atom beam even if threshold cannot be fully reached.

A central technical problem to solve is to achieve a sufficiently fast supply of new atoms. Any realistic description of this of course would require at least 2D modeling of the atomic motion. While this seems straightforward in principle, such models would be tedious to solve in practice and require large simulation efforts. However, we do not expect qualitative changes in the results overall.

Let us finally mention that intracavity Raman gain has been already observed and used for cooling in an experimental setup by Vuleti\'c and co-workers in Cesium \cite{Chan03}. Here the initial atomic temperature was quite high and there was no external lattice or trap present, but the results should give hope for current improved efforts \cite{Klinner06} starting from much colder sources.

\acknowledgments
We would like to thank P.\ Domokos, A.\ Kuhn, and J.\ Klinner for fruitful discussions.
This work was supported by the Austrian Science Foundation FWF, grants S1512 and P17709.

\end{document}